\documentclass[twocolumn]{aastex62} %linenumbers
\usepackage{amsmath}
\usepackage{longtable, booktabs, threeparttablex}
\graphicspath{{./}{figures/}}

\shorttitle{The Inner Kernel of the Classical Kuiper Belt}
\shortauthors{Siraj, Chyba, \& Tremaine}
\begin{document}

\title{The Inner Kernel of the Classical Kuiper Belt}

\email{siraj@princeton.edu}

\author{Amir Siraj}
\affil{Department of Astrophysical Sciences, Princeton University, 4 Ivy Lane, Princeton, NJ 08544, USA}

\author{Christopher F. Chyba}
\affil{Department of Astrophysical Sciences, Princeton University, 4 Ivy Lane, Princeton, NJ 08544, USA}
\affil{School of Public and International Affairs, Princeton University, 20 Prospect Lane, Princeton, NJ 08540, USA}

\author{Scott Tremaine}
\affil{School of Natural Sciences, Institute for Advanced Study, Princeton, NJ 08540, USA}

\begin{abstract}

The `kernel' of the classical Kuiper belt was discovered by Petit et al.\ (2011) as a visual overdensity of objects with low ecliptic inclinations and eccentricities at semimajor axes near $ \sim 44\mbox{\;AU}$. This raises the question -- are there other structures present in the classical Kuiper belt? If there are, clustering algorithms applied to orbits transformed into free elements may yield the best chance of discovery. Here, we derive barycentric free orbital elements for objects in the classical Kuiper belt, and use the Density-Based Spatial Clustering of Applications with Noise (DBSCAN) algorithm to identify a new structure, which we dub the inner kernel, located at $a \sim 43 \mathrm{\; AU}$ just inward of the kernel ($a \sim 44 \mathrm{\; AU}$), which we also recover. It is yet unclear whether the inner kernel is an extension of the kernel or a distinct structure. Forthcoming observations, including those by the Vera C. Rubin Observatory's Legacy Survey of Space and Time (LSST) may provide further evidence for the existence of this structure, and perhaps resolve the question of whether there are two distinct structures.

\end{abstract}

\keywords{Solar system -- Trans-Neptunian objects -- Kuiper belt}

\section{Introduction}
\cite{2011AJ....142..131P} discovered the `kernel' of the classical Kuiper belt as a visual overdensity 
 at semimajor axis $a \sim 44 \mathrm{\; AU}$ and eccentricity $e \sim 0.05$ in the classical Kuiper belt, observed in $e$ vs. $a$ and inclination $i$ vs.\ $a$ orbital 
element plots. In this \textit{Letter}, we ask the following question: can clustering algorithms identify additional structures in the Kuiper belt that have not yet been identified by eye?

Clustering algorithms like Density-Based Spatial Clustering of Applications with Noise (DBSCAN) have been used to identify structures in Gaia data \citep{2018A&A...618A..59C, 2019A&A...628A.123Z, 2021A&A...646A.104H, 2022A&A...664A.175P}, but they have not yet been applied to the Kuiper belt. The utility of these clustering algorithms will only increase as vast quantity of new data become available in the coming years, including those from the Vera C. Rubin Observatory's upcoming Legacy Survey of Space and Time (LSST).

When interrogating the architecture of the Kuiper belt, barycentric orbital elements should be used rather than heliocentric orbital elements, and the free or proper components of inclination and eccentricity should be used rather than the sum of the free and forced components, which is directly observed. Heliocentric orbital elements of Kuiper belt objects (KBOs) vary on a timescale comparable to Jupiter's orbital period as the Sun orbits the barycenter of the solar system, and the forced components of the eccentricity and inclination vary on the secular timescale. Thus barycentric free elements are better choices for identifying primordial structures.

In Section \ref{sec:formalism}, we describe the method we use to isolate the free components of eccentricity and inclination of KBOs. In Section \ref{sec:application}, we apply the procedure of Section \ref{sec:formalism} to the classical Kuiper belt, and we use a clustering algorithm on the $(a_{\rm bary}, e_{\rm free}, i_{\rm free})$ data to look for new structures in phase space in addition to the kernel. We recover the kernel, and we report the existence of a new structure at $\sim 43 \mathrm{\; AU}$, which we call the `inner kernel'.

\section{Formalism}
\label{sec:formalism}
The formalism we employ for computing the free inclinations and eccentricities of KBOs, described in this section, is based upon sections A.1 and A.2 of \cite{2022ApJS..259...54H}, chapter 8 of \cite{2002mcma.book.....M}, and chapter 7 of \cite{1999ssd..book.....M}. Averaging over both fast angles and assuming that all planets are on coplanar circular orbits, the Hamiltonian accounting for the planetary perturbation by the $j$-th planet is
\begin{equation}
\label{integral}
    \mathcal{H}_j = -\frac{\mathcal{G}m_j}{\pi^2} \int_{0}^{2 \pi} \sqrt{\frac{1 - \eta / 2}{r^2 + a_j^2}} K(\eta) \mathrm{d}M \; \; ,
\end{equation}
where $m_j$ is the mass of the $j$-th planet, $a_j$ is the semimajor axis of the $j$-th planet, $r$ is the barycentric distance of the KBO, $K(\cdot)$ is an elliptic integral, $M$ is the mean anomaly of the KBO, and
\begin{equation}
    \eta = \frac{4 a_j \sqrt{x^2 + y^2}}{r^2 + a_j^2 + 2 a_j \sqrt{x^2 + y^2}} \; \; ,
\end{equation}
where $x$ and $y$ are the barycentric coordinates of the KBO projected onto the orbital plane of the planet, which is taken to be equivalent to the ecliptic plane. The integral in Equation \ref{integral} is computed in a setup with the Sun, the giant planets, and the KBO initialized with its current barycentric orbital elements. 

The Hamiltonian accounting for the planetary perturbation by all four giant planets is
\begin{equation}
    \mathcal{H} = \sum_{j = 5}^{8} \mathcal{H}_j \; \; .
\end{equation}

We use $a$, $e$, and $I$ to denote the semimajor axis, eccentricity, and inclination of the KBO, respectively. The Delaunay variables are 
\begin{align*} 
L &=  \sqrt{\mathcal{G} M_\odot a} \\ 
G &=  \sqrt{\mathcal{G} M_\odot a (1 - e^2)} \\
H &= \sqrt{\mathcal{G} M_\odot a (1 - e^2)} \cos{I} \; \; ,
\end{align*}
and the conjugate angles are $M$, argument of perihelion $\omega$, and longitude of the ascending node $\Omega$, respectively. The longitude of perihelion is $\varpi\equiv\omega+\Omega$.

We use the $I-\Omega$ and $e-\varpi$ eigenfrequencies from \cite{1950USNAO..13...81B} for the solar system -- namely, $f_i$ where $1 \leq i \leq 8$ and $g_i$ where $1 \leq i \leq 10$, respectively. The associated eigenvectors are $I_{ij}$ and $e_{ij}$, and the associated phases are $\gamma_i$ and $\beta_i$. The extra two frequencies in the $e-\varpi$ solution are $g_9 = 2g_5 - g_6$ and $g_{10} = 2g_6 - g_5$ \citep{1950USNAO..13...81B, 1999ssd..book.....M}.

\subsection{Free Inclination}
The nodal precession rate contributed by each planet is
\begin{equation}
    B_j = \dot{\Omega}_j = \frac{\partial \mathcal{H}_j}{\partial H} \; \;.
\end{equation}
The components of the forcing poles attributable to planetary perturbations are\footnote{The notation differences here relative to \cite{1999ssd..book.....M} are as follows: $B_j$ and $A_j$, $\mu_i$ and $\nu_i$ all have opposite signs, but $B$ and $A$ have the same sign. The $i$ and $j$ indices always refer to secular modes and planets, respectively.} 
\begin{equation}
    p_{\text{forced}} = \sum_{i = 5}^{8} \frac{\mu_i}{B - f_i} \sin{(\gamma_i)} \; \; ,
\end{equation}
\begin{equation}
    q_{\text{forced}} = \sum_{i = 5}^{8} \frac{\mu_i}{B - f_i} \cos{(\gamma_i)} \; \; ,
\end{equation}
where
\begin{equation}
    \mu_i = \sum_{j = 5}^{8} B_j I_{ij} \; \;,
\end{equation}
and the total nodal precession rate is 
\begin{equation}
    B = \sum_{j = 5}^{8} B_j \; \; .
\end{equation}
The free inclination is 
\begin{equation}
    I_{\text{free}} = \sqrt{(p - p_{\text{forced}})^2 + (q - q_{\text{forced}})^2} \; \; ,
\end{equation}
where
\begin{equation}
    p = I \sin{\Omega} \; \;,
\end{equation}
\begin{equation}
    q = I \cos{\Omega} \; \; .
\end{equation}

\subsection{Free Eccentricity}
The pericenter precession rate contributed by each planet is
\begin{equation}
    A_j = \dot{\varpi}_j  = \dot{\omega}_j  + \dot{\Omega}_j = \frac{\partial \mathcal{H}_j}{\partial G} + \frac{\partial \mathcal{H}_j}{\partial H} \; \;,
\end{equation}

The components of the forcing poles attributable to planetary perturbations are 
\begin{equation}
    h_{\text{forced}} = \sum_{i = 5}^{10} \frac{\nu_i}{A - g_i} \sin{(\beta_i)} \; \; ,
\end{equation}

\begin{equation}
    k_{\text{forced}} = \sum_{i = 5}^{10} \frac{\nu_i}{A - g_i} \cos{(\beta_i)} \; \; ,
\end{equation}
where
\begin{equation}
    \nu_i = \sum_{j = 5}^{8} A_j e_{ij} \; \; ,
\end{equation}
and the total pericenter precession rate is 
\begin{equation}
    A = \sum_{j = 5}^{8} A_j \; \; .
\end{equation}
The free eccentricity is, 
\begin{equation}
    e_{\text{free}} = \sqrt{(h - h_{\text{forced}})^2 + (k - k_{\text{forced}})^2}
\end{equation}
where
\begin{equation}
    h = e \sin{\varpi} \; \;,
\end{equation}
\begin{equation}
    k = e \cos{\varpi} \; \; .
\end{equation}

\begin{longtable*}{@{}lllllllll@{}}
\caption{Calculated properties of each of the 1650 classical \newline multi-opposition KBOs in \cite{2024RNAAS...8...36V}. This table is published in its entirety in the machine-readable format. A portion is shown here for guidance regarding its form and content. Note that $q_{\rm forced} \equiv I_{\rm forced} \cos{(\Omega_{\rm forced})}$, $p_{\rm forced} \equiv I_{\rm forced} \sin{(\Omega_{\rm forced})}$, $k_{\rm forced} \equiv e_{\rm forced} \cos{(\varpi_{\rm forced})}$, and $h_{\rm forced} \equiv e_{\rm forced} \sin{(\varpi_{\rm forced})}$.} \\
\toprule
Name       & $I_{\rm free}  {[}\rm{deg}{]}$ & $e_{\rm free}$ & $I$ {[}\rm{deg}{]} & $e$ & $q_{\rm forced}$ [deg] & $p_{\rm forced}$ [deg] & $k_{\rm forced}$ & $h_{\rm forced}$\\* \midrule
\endfirsthead
\endhead
\bottomrule
\endfoot
\endlastfoot
\label{longtable}
1992 QB1  & 2.68328               & 0.06986                & 2.18547          & 0.06933 & 0.20021 & 1.78291 & -0.000509 & -0.000399             \\
1994 GV9  & 1.96995               & 0.06116                & 0.56387           & 0.06119 & 0.27615 & 1.81370 & -0.000037 & 0.000021              \\
1997 CR29  & 17.73925               & 0.21190                & 19.15098          & 0.21301 & -0.25623 & 1.59814 & 0.000408 & 0.001032              \\
... & ...               &...                & ...          & ...              \\
2022 HE1 & 2.38243               & 0.07369                & 0.76524          & 0.07421 & 0.60393 & 1.94745 & -0.000562 & -0.000458              \\
2022 QD261 & 6.17458               & 0.16292                & 4.57085          & 0.16295 & -0.05614 & 1.67800 & 0.000206 & 0.000247              \\
2022 QE261 & 7.39887               & 0.11694                & 7.44322          & 0.11703 & -0.08167 & 1.66813 & 0.000370 & 0.000487               \\* \bottomrule
\end{longtable*}

\section{Application \& Clustering Analysis}
\label{sec:application}

\begin{figure*}%[hptb]
 \centering
\includegraphics[width=\linewidth]{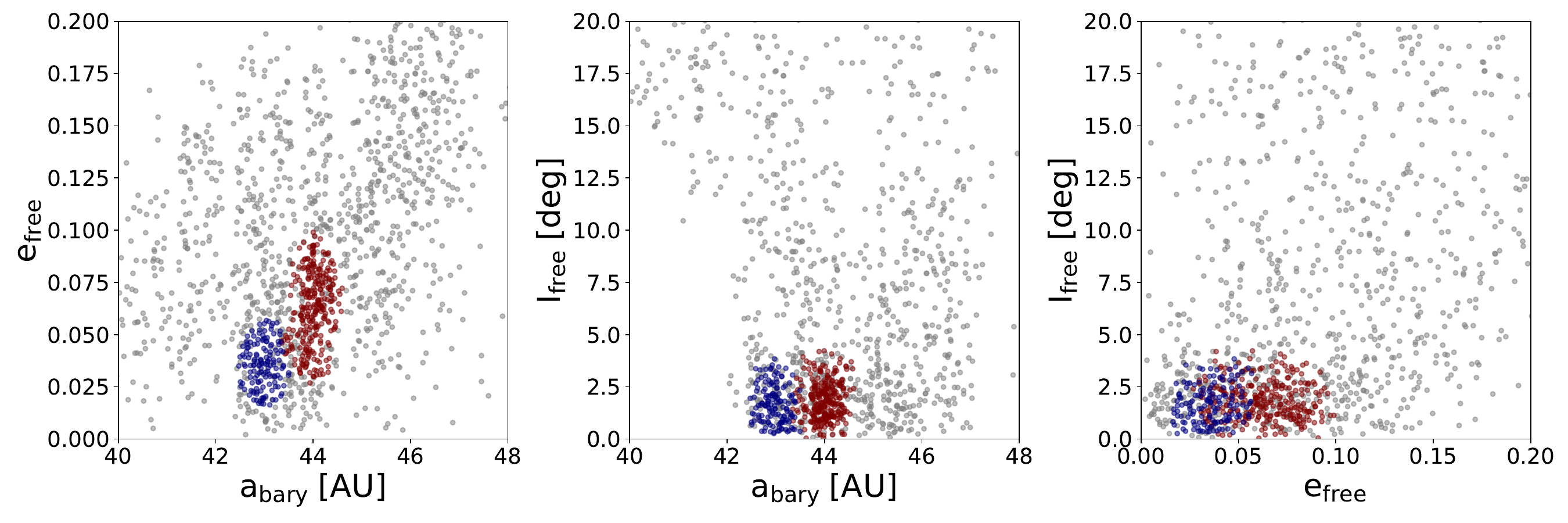}
\caption{DBSCAN results for $N_{min} = 50$ and $\epsilon = 0.15$. Kernel-like cluster in red, new cluster in blue, non-clustered classicals in gray.}
\label{fig:scatter}
\end{figure*}

\begin{figure*}%[hptb]
 \centering
\includegraphics[width=\linewidth]{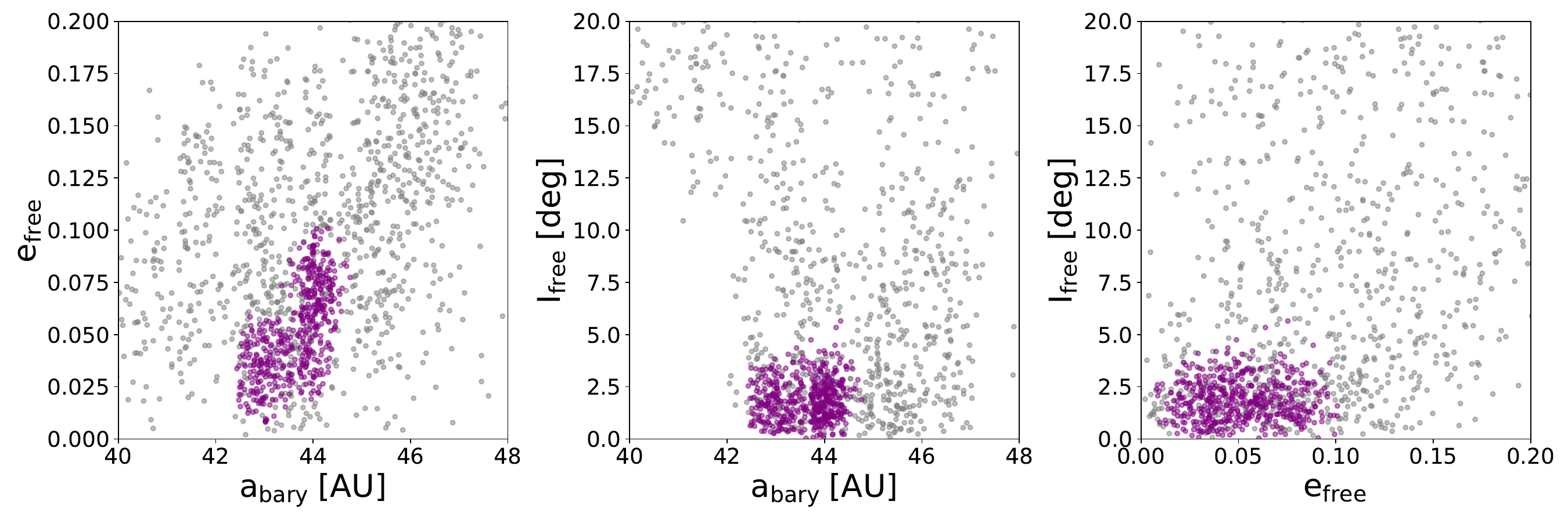}
\caption{DBSCAN results for $N_{min} = 50$ and $\epsilon = 0.16$, illustrating (relative to Figure \ref{fig:scatter}) that a slightly higher value of $\epsilon$ can lead to a combined kernel plus inner kernel cluster (shown here in purple). As in Figure \ref{fig:scatter}, the non-clustered classicals are shown in gray.}
\label{fig:scatter2}
\end{figure*}

\begin{figure*}%[hptb]
 \centering
\includegraphics[width=\linewidth]{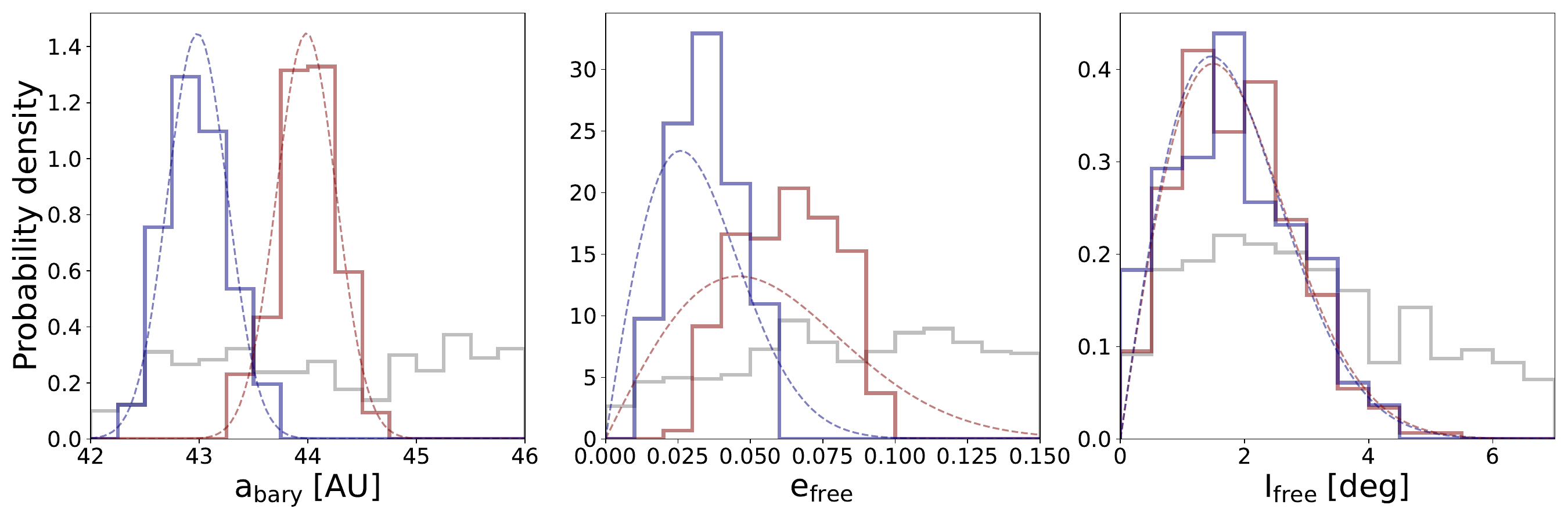}
\caption{Normalzed histograms of the $a_\text{bary}$, $e_{\text{free}}$, and $I_{\text{free}}$ distributions shown in Figure \ref{fig:scatter}. Dotted lines show best-fit Gaussian distribution for $a_\text{bary}$ and Rayleigh distributions for $e_{\text{free}}$ and $I_{\text{free}}$. These kernel-like and newly identified clusters have distributions of $a_\text{bary}$ described by Gaussians with means of $44.00 \mathrm{\; AU}$ and $42.98 \mathrm{\; AU}$ and standard deviations of $0.25 \mathrm{\; AU}$ and $0.24 \mathrm{\; AU}$, respectively. Rayleigh distributions with modes of $0.026$ and $0.046$ describe their respective distributions of $e_{\text{free}}$, and dispersions of $1.4^{\circ}$ and $1.3^{\circ}$ describe their respective distributions of $I_{\text{free}}$. The fit of the $e_{\rm free}$ distributions to the Rayleigh form is not good; fitting these distributions instead with Gaussians would yield means of $0.063$ and $0.036$, and standard deviations of $0.017$ and $0.011$, respectively.} Non-clustered classicals in gray.
\label{fig:hist}
\end{figure*}

\begin{figure*}%[hptb]
 \centering
\includegraphics[width=\linewidth]{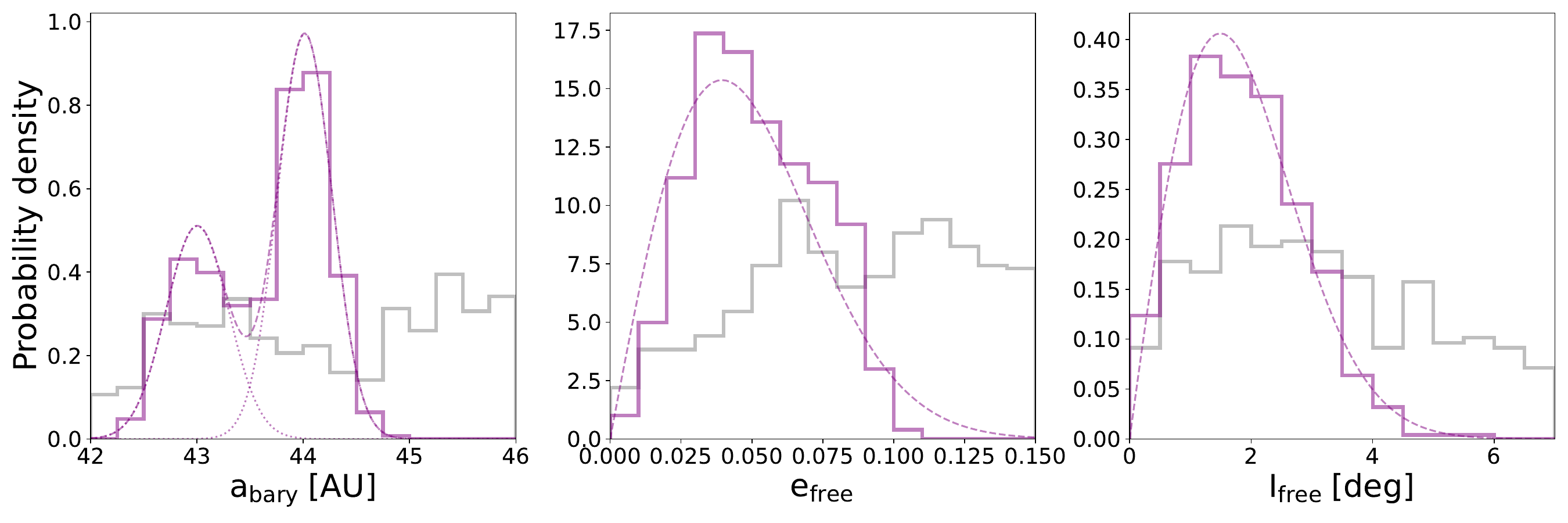}
\caption{Normalized histograms of the $a_\text{bary}$, $e_{\text{free}}$, and $I_{\text{free}}$ distributions shown in Figure \ref{fig:scatter2}. Dotted lines show best-fit bi-modal Gaussian distribution for $a_\text{bary}$ and Rayleigh distributions for $e_{\text{free}}$ and $I_{\text{free}}$. The distribution of $a_\text{bary}$ is described by two Gaussian distributions with means of $44.01\mbox{\;AU}$ and 
$43.00\mbox{\;AU}$, standard deviations of $0.26\mbox{\;AU}$ and $0.29\mbox{\;AU}$, and relative weights of 0.62 and 0.38, respectively. Rayleigh distributions with modes of $0.040$ and $1.5^{\circ}$ describe the distributions of $e_{\text{free}}$ and $I_{\text{free}}$, respectively. Fitting $e_{\rm free}$ instead with two Gaussians would yield means of $0.037$ and $0.072$, standard deviations of $0.013$ for each, and relative weights of $0.59$ and $0.41$, respectively. Non-clustered classicals in gray.}
\label{fig:hist2}
\end{figure*}

We calculate $I_{\text{free}}$ and $e_{\text{free}}$ for each of the 1650 multi-opposition KBOs in \cite{2024RNAAS...8...36V} labelled `classical' -- defined as non-resonant, non-scattering, and low eccentricity ($e \leq 0.24$). All values are listed in the machine-readable version of Table \ref{longtable}. Of the subset of these KBOs for which \cite{2022ApJS..259...54H} published $I_{\text{free}}$ values, our implementation produces $I_{\text{free}}$ values consistent with their results (defined as the difference being less than the \cite{2022ApJS..259...54H} value for the $I_{\text{free}}$ `range' for a given KBO) for $> 90\%$ of KBOs. The median and mean numerical differences in $I_{\rm free}$ between the \cite{2022ApJS..259...54H} values and those published here are $0.02^{\circ}$ and $0.06^{\circ}$, respectively. Improvements in the precision of the orbital elements for known KBOs, including data from additional oppositions in the three years between these two works, are a plausible source of the differences in $I_{\text{free}}$.

We use the \texttt{scikit-learn} implementation \citep{2011JMLR...12.2825P} of the density-based clustering algorithm DBSCAN \citep{1996kddm.conf..226E} to examine the architecture of the classical Kuiper belt. We use $a_\text{bary}$, $e_{\text{free}}$, and $I_{\text{free}}$, each re-scaled using \texttt{RobustScaler} to have a median of zero and a unit inter-quartile range \citep{2011JMLR...12.2825P, 2022A&A...664A.175P}. Specifically, \texttt{RobustScaler} subtracts the median and then divides by the inter-quartile range in order to re-scale each parameter. DBSCAN has two free parameters: $N_{min}$ and $\epsilon$. $N_{min}$ defines the minimum number of points required within distance $\epsilon$ of a point for it to be labeled a core point. 
Clusters are then defined as all points (core and non-core) that are reachable (an uninterrupted string of steps each $\leq \epsilon$) from a core point. In particular, one can keep walking in steps of $\epsilon$ and including new points in the cluster as long as those points also have $N_{min}$ points within a distance $\epsilon$. Border points (those with fewer than $N_{min}$ neighbors within a distance $\epsilon$) can be included but cannot be expanded from.

\begin{longtable}{@{}lll@{}}
\caption{\centering Properties of the inner kernel as identified \newline by DBSCAN when conditioned upon the detection of the \newline kernel as described in the text, across all $(N_{min}, \epsilon)$ pairs.} \\
\toprule
                                               & Inner Kernel        \\* \midrule
\endfirsthead
\endhead
\bottomrule
\endfoot
\endlastfoot
\label{table}
$a_\text{bary} \mbox{\;[AU]}$                       & $42.4 - 43.6$  \\
$e_{\text{free}}$                              & $0.01 - 0.06$   \\
Gaussian $\mu_{a,\rm bary} \mbox{\;[AU]}$          & $42.95 - 43.00$ \\
Gaussian $\sigma_{a,\rm bary} \mbox{\;[AU]}$       & $0.21 - 0.28$    \\
Rayleigh $\sigma_{e,\rm free}$      & $0.024 - 0.027$ \\
Gaussian $\mu_{e,\rm free}$      & $0.032 - 0.037$ \\
Gaussian $\sigma_{e,\rm free}$      & $0.009 - 0.011$ \\
Rayleigh $\sigma_{I,\rm free} \mbox{\;[deg]}$ & $1.3 - 1.5$ \\
Fraction of classicals                         & $7 - 10 \%$      \\
Fraction of cold classicals                    & $14 - 21 \%$   \\* \bottomrule
\end{longtable}

Different choices of $N_{min}$ and $\epsilon$ lead to different clustering results, so simply adopting a pair of ($N_{min}$, $\epsilon$) values and running the DBSCAN algorithm will yield a result that is inextricably tied to the choice of parameters. In order to mitigate this possibility, we adopt a few different reasonable values of $N_{min}$, and allow $\epsilon$ to vary freely for each. As a result, the results presented here depend minimally on the choice of clustering parameters. Specifically, we test $N_{min} = 25$, $50$, and $75$, allowing $\epsilon$ to vary freely for each in the range $10^{-3} - 1$.

For such a wide range of clustering parameters, a diverse set of possible clusters, at various scales, may arise. To yield a useful result, we perform our analysis in a conditional manner by asking a pointed, two-part question: 1. can DBSCAN identify a `kernel-like' cluster, and 2. if so, does it simultaneously identify any other clusters? The conditional analysis is necessary because on its own, running DBSCAN with varied $N_{min}$ and $\epsilon$ values will lead to a wide set of differing results. Conditionally anchoring any newly identified clusters to the necessitated existence of a known structure allows us to make a clear and specific statement with DBSCAN. By simply conditioning any results upon the existence of the kernel, we avoid having to make restrictive choices about $N_{min}$ and $\epsilon$ that would limit the usefulness of such an analysis.

The `kernel' refers to the subset of cold classical KBOs that have semimajor axes in the range $43.8 - 44.4 \mathrm{\; AU}$ and eccentricities in the range $0.03 - 0.08$ \citep{2011AJ....142..131P}. For the definition of the kernel only, we impose a requirement that $I_{\text{free}} < 4^{\circ}$ to restrict our sample to cold KBOs, as the kernel is defined as a subset of the cold classical Kuiper belt \citep{2021ARA&A..59..203G}. To address the first part of our question, we ask whether DBSCAN can identify a `kernel-like' cluster containing $> 90\%$ of the KBOs in the kernel as defined above without the overall ranges/widths in semimajor axis or eccentricity being dramatically larger (a factor of $> 3$) than the definition?\footnote{The use of slightly different values, like $> 95\%$ and $> 2$, for instance, does not change the qualitative results described here.} We find that the answer is yes, there is some range of values of $\epsilon$ that identify a kernel-like cluster for each value of $N_{min}$.\footnote{As for the ($N_{min}$, $\epsilon$) pairs that do not result in the identification of a kernel-like cluster: when $\epsilon$ is too large for a given value of $N_{min}$, DBSCAN typically identifies a cluster containing both the kernel and the inner kernel (which is no longer a `kernel-like cluster' because its extent in semimajor axis and eccentricity are much larger than that of the \citealt{2011AJ....142..131P} kernel); as $\epsilon$ grows, this cluster eventually encompasses the cold classical population, and at some point $\epsilon$ is so large that all of the objects are classified in a single cluster. When $\epsilon$ is too small for a given value of $N_{min}$, DBSCAN usually identifies a small population of the kernel and/or inner kernel, and as $\epsilon$ shrinks at some point it is so small that no clusters are identified whatsoever.} We then ask: when DBSCAN identifies a kernel-like cluster, does it simultaneously identify any other clusters? We find the answer to be yes: in all cases DBSCAN identifies one additional cluster at $a \sim 43 \mathrm{\; AU}$.

For this cluster (which we call the `inner kernel'), we list the overall ranges for various properties across all ($N_{min}$, $\epsilon$) pairs tested in Table \ref{table}. Figures \ref{fig:scatter} and \ref{fig:hist} illustrate the DBSCAN clustering results for nominal values of $N_{min}$ and $\epsilon$. We note that the absolute extent of the semimajor axis and eccentricity ranges found here for the kernel, $43.3 - 44.7 \mathrm{\; AU}$ and $0.03 - 0.10$ respectively, are similar to but slightly wider than the \cite{2011AJ....142..131P} definition of $43.8 - 44.4 \mathrm{\; AU}$ and $0.03 - 0.08$. This is likely attributable to the wide range of ($N_{min}$, $\epsilon$) pairs tested. Like \cite{2011AJ....142..131P}, we find that the kernel's inclination distribution is consistent with the cold classicals. We find the same to be true for the inner kernel.

In addition to being located at a smaller semimajor axis ($a \sim 43\mbox{\;AU}$) as compared to the kernel ($a \sim 44\mbox{\;AU}$), the inner kernel also has a colder free eccentricity distribution than the kernel, both in terms of the Rayleigh distribution dispersion that best describes it ($\sim 0.025$) as well as the overall range $(0.01 - 0.06)$. As a result, the inner kernel may provide stricter constraints than the kernel with respect to the degree of dynamical heating that the Kuiper belt could have plausibly sustained during the formation and evolution of the solar system. Perihelia of both the kernel and the inner kernel are safely above the stability boundary ($q \sim 37\mbox{ AU}$, see Figure 5 of \citealt{2021ARA&A..59..203G}). In addition, we calculated the maximum eccentricity for each object by summing the magnitudes of the free and forced eccentricities, and confirmed that both the kernel and the inner kernel are bounded by $q \gtrsim 40\mbox{ AU}$, which is distinctly beyond the stability boundary.

The discovery of the inner kernel, described here, was achieved by using DBSCAN in a conditional manner -- requiring that DBSCAN recover the kernel, and asking whether in those cases it simultaneously identified any other cluster(s). However, removing this condition, it is unclear whether the inner kernel and the kernel are a single, combined structure, or actually two distinct ones. Slightly increasing the $\epsilon$ value of an ($N_{min}$, $\epsilon$) pair that produced a distinct kernel and inner kernel can lead to the merging of the two structures, as illustrated in Figures \ref{fig:scatter2} and \ref{fig:hist2}. This is a particularly relevant as the apparent deficit of KBOs at semimajor axis values between the kernel and the inner kernel may be associated with the presence of the 7:4 mean-motion resonance with Neptune at $43.7\mbox{\;AU}$. As a result, there are two alternative explanations that we cannot distinguish between: either the kernel is significantly larger than previously thought, or there is an additional distinct structure in the cold classical Kuiper belt. In either case, the inner kernel, as described here, is the additional component. One suggestive piece of evidence that the inner kernel is distinct is that its eccentricity distribution is quite different from that of the kernel.

It is not obvious how the inner kernel was formed. Jumpy migration of Neptune is a possible explanation for the kernel that may also apply to the inner kernel \citep{2015AJ....150...68N}. The possibility that the inner kernel is a collisional family is somewhat disfavored by its relatively narrow extent in semimajor axis (see the discussion in \citealt{2015AJ....150...68N} regarding both \citealt{2002ApJ...573L..65C} and \citealt{2011ApJ...733...40M}).
\newline

%\newpage
\section*{Acknowledgements}

We thank the referee, Jean-Marc Petit, for thoughtful comments that improved the paper. We are pleased to acknowledge that the work reported in this paper was substantially performed using the Princeton Research Computing resources at Princeton University which is consortium of groups led by the Princeton Institute for Computational Science and Engineering (PICSciE) and Office of Information Technology's Research Computing.

%\newpage

%\appendix

\newpage

%\newpage
\bibliography{bib}{}
\bibliographystyle{aasjournal}

\end{document}